\documentclass[doublecol]{epl2}
\usepackage[latin1]{inputenc}
\usepackage{graphicx}
\usepackage{amssymb,amsmath}

\newcommand{\la}{\left\langle}
\newcommand{\ra}{\right\rangle}
\newcommand{\be}{\begin{equation}}
\newcommand{\ee}{\end{equation}}
\newcommand{\bse}{\begin{subequations}}
\newcommand{\ese}{\end{subequations}}
\newcommand{\bea}{\begin{eqnarray}}
\newcommand{\eea}{\end{eqnarray}}
\newcommand{\ba}{\begin{array}}
\newcommand{\ea}{\end{array}} 
	
\title{Macroscopic  arrow of time from multiscale perspectives}

\author{Mahendra K. Verma\inst{1}}
\institute
{ \inst{1} Department of Physics, Indian Institute of Technology, Kanpur 208016, India
}

\abstract {Fundamental laws of physics are symmetric under time reversal ($T$) symmetry, but the $T$ symmetry is strongly broken in the macroscopic world.  In   this Perspective,  I review $T$ symmetry breaking frameworks: \textit{second law of thermodynamics, multiscale energy transfer}, and \textit{open systems}. In driven dissipative nonequilibrium systems, including turbulence, the multiscale energy flux from large scales to small scales  helps determine the arrow of time.  In addition, open systems are often irreversible due to particle and energy exchanges between the system and the environment. Causality is another important factor that breaks the $T$ symmetry. 
}

\begin{document}
\maketitle

{\bf Introduction.}-- Fundamental laws of classical and quantum physics (except those involving weak nuclear force, which this {Perspective} ignores) are symmetric under time reversal, denoted by $T$~\cite{Kane:book:Particle}.   Suppose that {\bf x}($t$)  represents the classical trajectory of a particle. If we reverse the velocity of the particle at time $t_f$, then, according to $T$ symmetry, the particle will retrace the original trajectory {\bf x}($t$). An equivalent way is to film the original motion. If this movie played backward depicts the system evolution with the reversed velocity of the final state, then the system is reversible. Microscopic systems experiencing gravity and electromagnetic forces exhibit the above behaviour. However,  macroscopic systems have a definite direction of time~\cite{Boltzmann:book_chapter,Lebowitz:PT1993,Lebowitz:PA1993}. For example, a tree grows and then dies; pieces of a broken egg do not come together to form the original egg.   This apparent inconsistency of $T$ symmetry at the microscopic and macroscopic scales is called the \textit{arrow of time problem}~{ \cite{Carroll:book:Time,Halliwell:book,Mackey:book:Time,Davies:book:Time}}. The present {   Perspective} deals with this issue.

As pointed out by Boltzmann~\cite{Boltzmann:book_chapter,Lebowitz:PT1993,Lebowitz:PA1993,Carroll:book:Time,Halliwell:book,Gibbings:book,Landau:book:StatMech},  \textit{the second law of thermodynamics} breaks the   $T$ symmetry strongly in  macroscopic systems because the entropy of these systems always increase.   Note, however, that the second law  is rooted in thermodynamics, and it is not clear whether it could be \textit{convincingly} generalized to more complex systems such as earthquakes~\cite{Fowler:book}, turbulence~\cite{Lesieur:book:Turbulence},  human body, etc.  

For turbulent flows, Verma~\cite{Verma:EPJB2019} argued that  \textit{multiscale energy transfer} (or  \textit{energy flux} or \textit{energy cascade}) can break  $T$ symmetry, or provide an arrow of time. In a three-dimensional (3D) driven turbulence, the energy cascades from large scales to small scales~\cite{Lesieur:book:Turbulence}. In time-reversed movie of the flow, the energy will flow from small scales to large scales, which is absurd. Hence, we can use the energy flux as a diagnostic to determine the arrow of time in turbulence.  Similarly, the energy transfers  dictate the arrow of time in   other driven dissipative systems, e.g., earthquakes, fragmentation, and crack propagation, where  the energy flows from large scales to small scales~\cite{Verma:EPJB2019}. Some researchers~\cite{Jucha:PRL2014,Xu:PNAS2014} quantified time irreversibility in a turbulent flow using the deviations in the forward and reversed trajectories of Lagrangian particles.


The $T$ symmetry is also  broken in open systems because of  force, energy, and particle exchanges between the system and environment. For example, a periodically-forced oscillator and an electromagnetically-driven molecule are irreversible.  Chemical reactions (e.g., combustion), Earth's atmosphere (driven by the Sun), Earthquakes (driven by plate techtonics),  and astrophysical turbulence (driven by supernovas, jets, and winds) are prime examples of irreversible open systems. Causality, an important property of an open system, too breaks $T$ symmetry at the macroscopic level.   Note that entropy is not well defined for some open systems, e.g., earthquakes.  In addition, many molecular collisions  are irreversible due to the complex  interactions between the electrons of the atoms~\cite{Kotz:book}. 

In this {   Perspective} we briefly describe various { frameworks} that break $T$ symmetry at  macroscopic level.    For clarity, brevity, and focus,  I avoid  discussions on entropies of  information and computation~\cite{Shannon:BELL1948,Thurner:book},  black hole entropy~\cite{Giddings:PT2013}, Landauer's principle~\cite{Landauer:IBM2000}, and cosmological  and psychological arrows of time~\cite{Carroll:book:Time,Halliwell:book}. 

\begin{center}
	\begin{figure}[t]
		\onefigure[height=!, width=7cm]{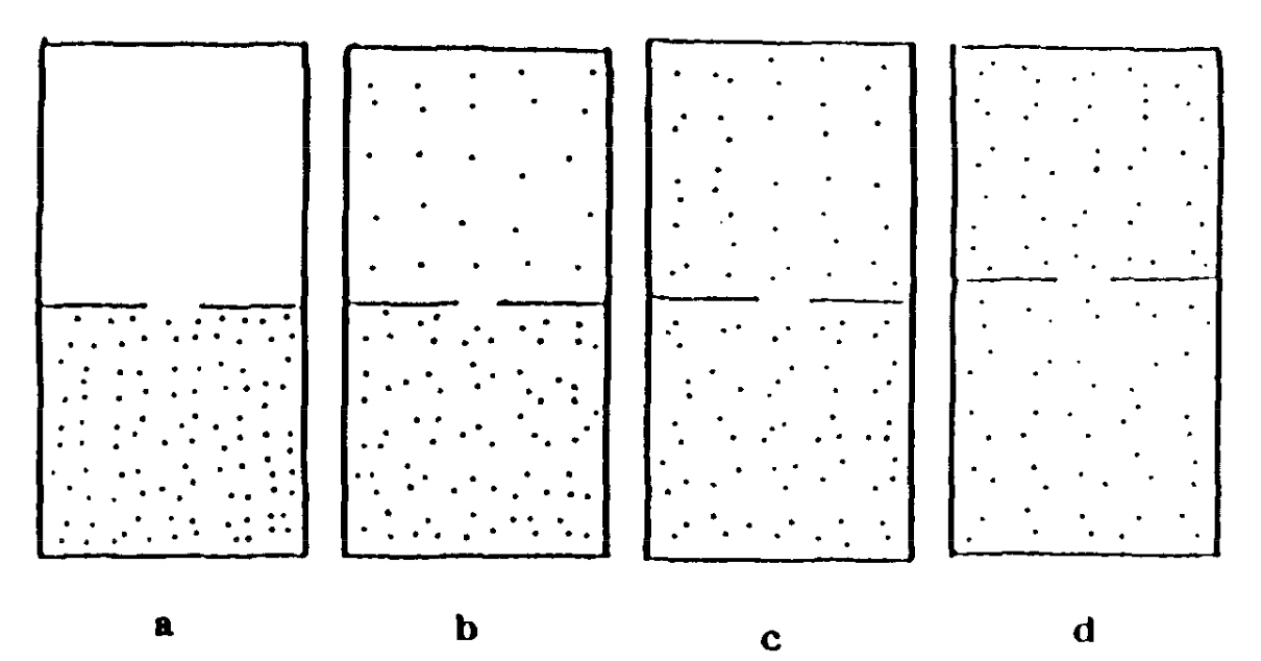}
		\caption{Successive evolution of the gas confined in the lower half of the box (a) to the  equilibrium state (d). Taken from Lebowitz~\cite{Lebowitz:PA1993}. Reprinted with permission from Elsevier. } 
		\label{fig:Lebowitz}
	\end{figure}
\end{center}

\vspace{0.3cm} 
{\bf Second law of thermodynamics.}--  The  second law of thermodynamics is most often employed to break the $T$ symmetry macroscopically~\cite{Boltzmann:book_chapter,Gibbings:book,Landau:book:StatMech,Carroll:book:Time,Halliwell:book}.  Lebowitz~\cite{Lebowitz:PA1993,Lebowitz:PT1993} presents Boltzmann's arguments as follows. Figure~\ref{fig:Lebowitz} illustrates four snapshots of the gas  expansion after the separator between the two halves is opened. The gas particles evolve via the following equations of motion: $m {\bf \dot{x}_\alpha = {p}_\alpha}$ and  ${\bf \dot{p}}_\alpha = -\nabla_\alpha \sum_\beta V_{\alpha,\beta}$, where $\alpha, \beta$ are particle labels; ${\bf x}_\alpha$ and  ${\bf p}_\alpha$ are respectively  the position and momentum of particle $\alpha$; and $V_{\alpha,\beta}$ is the inter-particle potential between particles $\alpha$ and $\beta$. All the particles have mass $m$.

A system configuration is represented by a point in the available phase space volume $\sum_\alpha \int d{\bf p}_\alpha d{\bf q}_\alpha$. Boltzmann~\cite{Boltzmann:book_chapter} constructs a macrostate by combining nearby configurations,  { which includes all  permutations of particles in a snapshot of Fig.~1~\cite{Boltzmann:book_chapter,Lebowitz:PA1993}.} We denote the macrostates corresponding to Fig.~\ref{fig:Lebowitz}(a,b,c,d) using  $\Gamma_a$, $\Gamma_b$, $\Gamma_c$, and $\Gamma_d$ respectively. Boltzmann derived the entropy of an ideal gas under equilibrium as~\cite{Boltzmann:book_chapter,Gibbings:book,Landau:book:StatMech}
\be 
S = {  k_B \log W} = 
N k_B \left[ \log \frac{V}{N} + \frac{3}{2} \log \frac{4\pi m E}{3Nh^2}  + \frac{5}{2} \right],
\label{eq:S_thermo_gas}
\ee
where {  $W$ is the number of microstates};  $E, N, V$ are respectively the energy, number of particles, and volume of the system; $h, k_B$ are respectively the Planck and Boltzmann constants; and $m$ is the mass of each particle. 
The above entropy,  referred to as the \textit{Boltzmann entropy},  is same as  the thermodynamic entropy proposed by Clausius, and it always increases with time until the system reaches an equilibrium~\cite{Landau:book:StatMech}.  A cautionary remark is in order. Equation~(\ref{eq:S_thermo_gas}) applies to an ideal gas under equilibrium, which may not be the case for Fig.~\ref{fig:Lebowitz}(b,c). However, we use the above entropy formula assuming quasi-equilibrium evolution.

As argued by Boltzmann~\cite{Boltzmann:book_chapter}, most configurations of Fig.~1(c) evolve to Fig.~1(d), which is the  \textit{equilibrium} state of the system.  This process is called \textit{thermalization}.  Note, however, that a \textit{precise} time reversal (exact ${\bf p}_\alpha \to -{\bf p}_\alpha$) of a configuration of Fig.~\ref{fig:Lebowitz}(c) will evolve to   Fig.~\ref{fig:Lebowitz}(b) and then to Fig.~\ref{fig:Lebowitz}(a). However, this is an \textit{atypical} event.  A small error in the reversed velocity  yields a very different result. For this case, a configuration of Fig.~\ref{fig:Lebowitz}(c) will evolve towards Fig.~\ref{fig:Lebowitz}(b) for a short while, and then it turns around and  evolves towards  Fig.~\ref{fig:Lebowitz}(c)  and then to Fig.~\ref{fig:Lebowitz}(d).  { Chaos theory indicates that a small error in the initial condition can get amplified in a nonlinear  system~\cite{Strogatz:book}. This feature plays an important role in the above randomization process~\cite{Mackey:book:Time}. These factors lead to an increase in the entropy of a thermodynamic system.  }

This theory of Boltzmann is a major advance in modelling macroscopic world.

\vspace{0.3cm}
{\bf Asymmetric energy transfers.}-- A fluid, which consists of atoms and molecules, has hydrodynamic and thermodynamic components. For Earth's atmosphere, the winds are the hydrodynamic component, whereas the quasi-equilibrium fluid at the centimeter scale is the thermodynamic component. Prigogine~\cite{Prigogine:Science1978} argued  that ``nonequilibrium may become a source of order and that irreversible processes may lead to a new type of dynamic states of matter called \textit{ dissipative structures}", and  pointed out that  thermodynamics may require some revision for describing  hydrodynamic structures.  Let us illustrate this aspect using Euler equation~{ \cite{Lesieur:book:Turbulence,Morrison:RMP1994}}:
\be
\frac{\partial {\bf u}}{\partial t} + {\bf u \cdot \nabla u} = -\nabla p ,
\label{eq:Euler}
\ee
where ${\bf u}, p$ are  the velocity and pressure fields respectively. We assume the fluid to be incompressible ($\nabla \cdot  {\bf u} = 0$).  

{ Equation~(\ref{eq:Euler}) is symmetric under time reversal operation: ${\bf u} \rightarrow {\bf u'} = -{\bf u}$ and $t \rightarrow t_r = t_f -t$, where $t_f$ is the final time of the evolution.  Here, the subscripts $r,f$ represent \textit{reversed} and \textit{forward}, respectively. The system evolves from $t=0$ to $t=t_f$ during the forward motion, and from $t_r =t_f$ to $t_r=0$ during the reversed motion\footnote{{ Often quoted transformation $t \rightarrow t_r = -t$ fails to incorporate the initial and final states properly. However, the transformation $t \rightarrow t_r  = t_f -t$ clearly indicates the initial and final states,  thus  contrasting the forward and backward motion \cite{Verma:book:Mechanics}.}}.  On exact time reversal, a fluid parcel would revert back to the original position, similar to the reversals of particles under time reversal. However, small perturbation (in experiments or numerical simulations) leads the fluid parcel to follow its natural evolution with forward energy flux in 3D (to be discussed in the following).  This is similar to a continual increase in entropy in a gas of Fig.~\ref{fig:Lebowitz} even after velocity reversals of the constituent particles.}

Note that the Boltzmann entropy of an Euler flow remains constant because of the zero viscosity and zero heat dissipation~\cite{Landau:book:StatMech}. However, past numerical simulations, e.g., Cichowlas {\em et al.}~\cite{Cichowlas:PRL2005},  show that the correlations or order of the flow changes with time.  Constancy of the Boltzmann entropy in  Euler flow forces us to go beyond  classical thermodynamic framework for determining the arrow of time in the Euler turbulence~\cite{Verma:PRE2024}. Fortunately, the multiscale energy flux in Euler turbulence helps in this regard~\cite{Verma:EPJB2019,Verma:PRF2022,Jucha:PRL2014}. 

In a 3D Euler flow with an ordered initial condition, the energy flows from large scales to small scales~\cite{Cichowlas:PRL2005}. The transferred energy gets equi-distributed among the small-scale modes, who reach a quasi-equilibrium state~\cite{Cichowlas:PRL2005}. Note that in a time-reversed Euler flow, the energy will flow backward (from small scales to large scales), which is absurd and unphysical.  Hence, the energy flux helps us determine the arrow of time for Euler turbulence.  

{Asymptotically, Euler turbulence thermalizes to an equilibrium configuration, with all the Fourier modes having equal energy. Note that the equilibrium state does not revert to the original configuration; it is reasonable to assume that the Poincar\'{e} recurrence time is too large for large degrees of freedom in Euler turbulence.  Thermalization in Euler turbulence appears to have certain similarities with those in quantum systems~\cite{Rigol:Nature2008}; this connection needs to be examined in detail. 
}

Euler  turbulence illustrates another important property of \textit{time}.  In  nonequilibrium Euler turbulence, the large-scales components  move forward in time (due to the positive energy flux), but the small-scale components are in equilibrium with frozen time. This example indicates that in a macroscopic system, time at different scales flow at different rates. Note that various parts in a human body  (e.g., cell and heart) evolve at different rates.

Euler equation has  zero viscosity, but realistic flows have nonzero kinematic viscosity ($\nu$). The following incompressible Navier-Stokes (NS) equation describes such flows quite well~\cite{Lesieur:book:Turbulence}:
\be
\frac{\partial {\bf u}}{\partial t} + {\bf u \cdot \nabla u} = -\nabla p  + \nu \nabla^2 {\bf u}+{\bf F}_\mathrm{ext},
\label{eq:NS}
\ee
where $\nu$ is the kinematic viscosity, and  ${\bf F}_\mathrm{ext}$ is the external force, which is active at large scales in Kolmogorov's theory. 
We assume the density of the fluid to be unity, which implies that $\nabla \cdot  {\bf u} = 0$~\cite{Lesieur:book:Turbulence}. { Clearly, the viscous term $\nu \nabla^2 {\bf u}$ breaks the time-reversal symmetry. In addition, time-asymmetric ${\bf F}_\mathrm{ext}$ too will break the $T$ symmetry (to be discussed along with \textit{open systems}). Note that the viscous term  converts the coherent hydrodynamic energy  to the thermal energy, which is an irreversible process.}

 The energy flux, which is significant at large and intermediate scales, too  determines the  arrow of time for this system~\cite{Verma:EPJB2019}. In particular, Kolmogorov~\cite{Kolmogorov:DANS1941Dissipation,Lesieur:book:Turbulence} derived the following  third-order structure function for a homogenous, isotropic, and steady turbulent flow:
\be
S_3(r) = \la ([{\bf u(x+r)-u(x)}] \cdot{\bf \hat{r}})^3 \ra = -\frac{4}{5} \epsilon_u r,
\label{eq:S3}
\ee 
where ${\bf u(x)}$ and ${\bf u(x+r)}$ are the velocities at the positions at ${\bf x}$ and ${\bf x+r}$ respectively;  ${\bf \hat{r}}$ is the unit vector along ${\bf r}$; and $\epsilon_u$ is the energy flux. A restriction is that $r$ should be much smaller than the system size, but much larger than the viscous scale. Note that $\epsilon_u$ is positive, hence $S_3(r)$ is negative. However, $S_3(r)$  changes sign under the time-reversal operation, ${\bf u \to u' = -u}$. That is, for a snapshot of a time-reversed flow, 
\be
S'_3(r) = \la ([{\bf u'(x+r)-u'(x)}] \cdot{\bf \hat{r}})^3 \ra = \frac{4}{5} \epsilon_u r > 0.
\ee
Therefore, the  time-reversed flow  exhibits inverse energy cascade, which is physically unrealizable.  Hence, for Navier-Stokes equation too, the energy flux differentiates the forward and time-reversed flows and breaks  the $T$ symmetry. 

Similar to the  ideal gas of Fig.~\ref{fig:Lebowitz}, under exact time reversal, Euler flow [Eq.~(\ref{eq:Euler})] will return to its initial state.  However, with viscosity and/or numerical and experimental errors during the velocity reversal,  the time-reversed flow will follow its natural  direction with positive energy flux after a transient with negative energy flux (to be illustrated below using the \textit{shell model}).    This is similar to the gas expansion illustrated in Fig.~\ref{fig:Lebowitz}. As described earlier, on velocity reversal of the gas, the system moves towards Fig.~1(b), but it quickly turns around and evolves towards Fig.~1(d). These observations indicate that the forward energy flux of a 3D turbulent flow  is  robust under perturbation.

\begin{center}
	\begin{figure}[t]
		\onefigure[height=!, width=8cm]{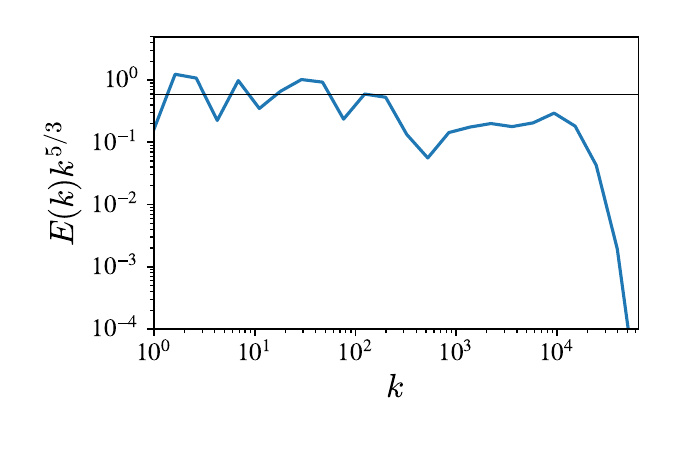}
		\caption{Plot of the normalized energy spectrum $E(k) k^{5/3}$ vs.~$k$ for a fully developed turbulent flow in a shell model. We observe that $E(k) k^{5/3} \approx 0.6$, a constant in the inertial range. } 
		\label{fig:ek}
	\end{figure}
\end{center}

\begin{center}
	\begin{figure}[t]
		\onefigure[height=!, width=8cm]{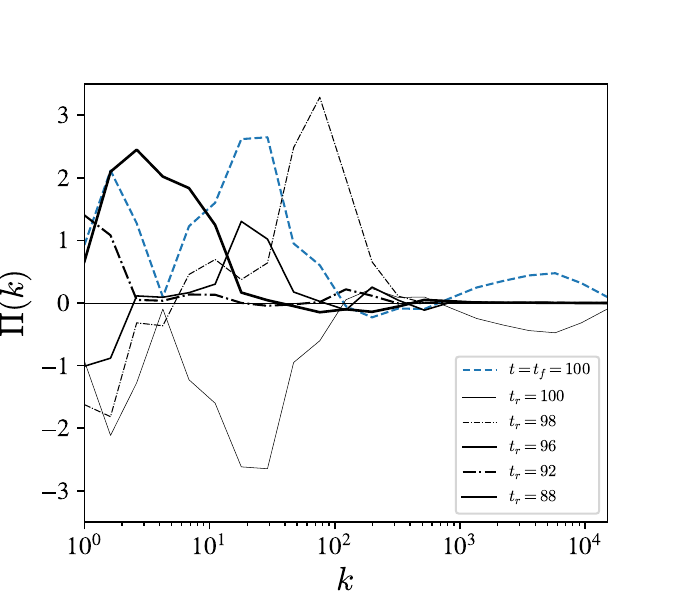}
		\caption{Plots of the energy flux $\Pi(k)$ vs.~$k$ for the forward and time-reversed  flows. The dashed blue curve represents  $\Pi(k)$ at $t=t_f=100$, at which point we reverse $\{u_n\}$ that leads to inverse $\Pi(k)$ and evolve the system backward from $t_r = 100$ to 88.  Note that  $\Pi(k)$ crosses over from negative to positive  as $t_r$ crosses 92.} 
		\label{fig:flux}
	\end{figure}
\end{center}

We demonstrate the robustness of  energy flux using a quick turbulence simulation of a shell model with 30 shells and  $\nu = 10^{-6}$~\cite{Gledzer:DANS1973,Biferale:ARFM2003}.  The initial condition of the shell model is chosen is such a way that the energy flux at the large and intermediate scales is one unit. We time advance the decaying shell model using RK4 scheme with time-step $dt = 10^{-5}$.  Since $\nu \to 0$, the  shell model loses insignificant amount of energy, and reaches a quasi steady-state.  We compute the energy spectrum $E(k)$ and energy flux $\Pi(k)$ for a steady-state snapshot at $t=t_f = 100$ non-dimensional unit.  For this snapshot, as shown in Fig.~\ref{fig:ek}, $E(k) k^{5/3} \approx $ const, implying that  $E(k) \propto k^{-5/3}$.  In Fig.~\ref{fig:flux}, the dashed-blue curve illustrates  $\Pi(k)$, which is positive. Note that the energy flux for a snapshot has large fluctuations, which are smoothened out on averaging.  In this {   Perspective}, we compare the temporal evolution of $\Pi(k)$, hence we will have to live with these large fluctuations.

At $t=t_f=100$, we perform time reversal operation, $\{u_n\} \to \{-u_n\}$, which leads to an exact opposite energy flux (the thinnest black curve of Fig.~\ref{fig:flux}).  We { evolve  the shell model using the same parameters  (with the reversed velocity field) in reversed time from $t_r=100$ to $t_r=88$.}  We observe that the energy flux gradually changes sign (from negative  to positive), as illustrated in Fig.~\ref{fig:flux}.  Finally, at $t \approx  88$,  the energy flux becomes positive,  which is its natural direction.   We plan to demonstrate the above phenomena using a high-resolution   simulation of NS equation, which  will be much more expensive than the shell model.

{ Thus, asymmetric energy transfer is  useful diagnostic (apart from entropy)  for determining the arrow of time in a turbulent flow.} This is particularly useful because the Boltzmann entropy has certain ambiguities in hydrodynamics~\cite{Prigogine:Science1978,Kondepudi:book:Thermo,Landau:book:StatMech}. { Recently, Verma {\em et al.} \cite{Verma:PRF2022,Verma:PRE2024} formulated \textit{hydrodynamic entropy} that appears to capture the multiscale entropy of turbulence and other related systems (e.g., coarsening in Time-dependent Ginzburg-Landau equation). This entropy is defined as $S_H = - \sum_{\bf k} p_{\bf k} \log_2 p_{\bf k}$, where $p_{\bf k} = E({\bf k})/ [\sum_{\bf k}E({\bf k}) ]$ with $E({\bf k})$ as the modal energy for wavenumber \textbf{k}.  For white noise (delta-correlated) corresponding to the equilibrium state, $S_H = \log_2 M$ with $M$ as the degrees of freedom,  is the  maximum entropy. In contrast, $S_H $ for  a turbulent flow is between 3 and 4, much less than the maximum entropy~\cite{Verma:PRE2024}.  Refer to \cite{Verma:PRF2022,Verma:PRE2024} for more details. }

Turbulence is observed everywhere, e.g., in atmospheres~\cite{Salby:book}, in astrophysical objects~\cite{Zeldovich:book:Magneticfield}, in plasma turbulence~\cite{Matthaeus:APJ2020}, and in engineering flows~\cite{Lesieur:book:Turbulence}.  Hence we can employ energy flux to define arrow of time in such flows. Interestingly,  the energy flux can  also be defined for  quantum turbulence and Gross-Pitaevskii equation~\cite{Barenghi:PNAS2014},  coarsening~\cite{Verma:PRE2023_coarsening}, molecular dynamics~\cite{Wani:Arxiv2023}, earthquakes~\cite{Fowler:book}, fragmentation, crack propagation,  and animal body (driven by food intake).  The aforementioned  systems involve complex and varied physics, but a   generic feature among them is  a multiscale energy flux, which is induced by large-scale forcing and small-scale dissipation.  

Among the above driven nonequilibrium, quantum systems, Euler equation, and Gross-Pitaveskii equation, are conservative or Hamiltonian.   In multiscale Hamiltonian systems driven at large scales, the large-scale coherent hydrodynamic energy cascades to small scales, where the small-scale hydrodynamic energy is converted to incoherent thermal energy.  This  conversion of hydrodynamic energy from coherent structures to incoherent thermal energy can be treated as \textit{effective dissipation}~\cite{Verma:EPJB2019}.   In quantum turbulence,  phonons generated at small scales act as energy sink, and help generate an energy flux~\cite{Barenghi:PNAS2014}.

An often-quoted irreversible process---breakage of an egg---too involves multiscale energy transfers.  When a cook strikes an egg with a spoon, cracks  formed on the egg shell  propagate through the surface.  At some point of time,  the shell breaks and sound waves are released.  Formation and propagation of cracks involves irreversible condensed-matter  processes~\cite{Kittel:book:Solid}. But, the forward energy flux provides a simple $T$ breaking mechanism for  this system.

{ Two-dimensional turbulence has peculiar properties. When we start the flow with a vortex comparable to the box size, the energy flows from large scales to the small scales~\cite{Wani:PLA2024}, similar to 3D hydrodynamic turbulence. However, when 2D turbulence is forced at  intermediate scales,  the kinetic energy exhibits an inverse cascade, whereas  the enstrophy shows a forward cascade~\cite{Kraichnan:JFM1973}. For the latter  2D flow, the negative energy flux is the natural direction that determines the arrow of time; the corresponding time-reversed flow  will exhibit positive energy flux, which is not realizable in these systems~\cite{Verma:EPJB2019}.  Thus, time's arrow of a system is dictated by the natural direction of energy transfer in that system. 
}

{  The forward energy flux is an inherent property of 3D turbulence, with the scale separation between  the external force and viscous dissipation playing a critical role. The energy injected by external force spreads out to intermediate wavenumbers and then to the dissipative range, where the hydrodynamic energy is dissipated. This energy spread  appears to occur in most turbulent flows, except in 2D turbulence forced at intermediate scales where the energy cascades backward. However, the latter anomaly is possibly related to 2D being a special dimension, as stated in the \textit{Mermin-Wagner theorem}~\cite{Chaikin:book}. These trends in energy transfers hold even in isolated systems, with the \textit{Poincar\'{e} recurrence theorem} yielding a very minute probability for the time reversal.  The spread of  energy and entropy to available spaces  appears to  have  similar inherent dynamics. For example, in Fig.~\ref{fig:Lebowitz}, the particles spread out to all available volume. We plan to explore these connections in the future. 
}

In terms of particle picture,   irreversibility in a turbulent flow can be quantified using   the difference between the forward and backward dispersion of Lagrangian particles~\cite{Jucha:PRL2014,Xu:PNAS2014}. But, unlike energy transfers, particle-based irreversibility measure is not easily generalizable to nonequilibrium systems without particles, e.g., earthquakes, human body, and quantum turbulence.

Unlike entropy, which is defined for Hamiltonian systems, energy flux can provide time's arrow   for both energy-conserving and dissipative systems.  These  features make multiscale energy transfer an important  ingredient for  breaking the $T$ symmetry. Next, we discuss how arrow of time comes out in open systems.

\vspace{0.3cm} 
{\bf Open Systems.}-- A simple example of an open system  is a periodically-forced linear oscillator, whose equation is
\be
m \ddot{x} = - m \omega_0^2 x + F_0 \sin(\omega t),
\label{eq:forced_osc}
\ee
where $m$ and $ \omega_0$ are respectively the mass and natural frequency of the oscillator; and $F_0$ and $\omega$ are respectively the amplitude and frequency of the forcing. The solution $x(t)$ of Eq.~(\ref{eq:forced_osc}) is not $T$ symmetric due to the time-dependent external forcing. Many macroscopic systems in the world are open and driven. For example, hydrodynamic turbulence is often forced at large scales;  many quantum systems are driven by lasers and interactions; Earth's atmosphere is driven by the Sun and large-scale winds;  and galactic turbulence is driven by supernovae, jets, and winds. The external time-dependent forces, along with  entropy and  multiscale energy fluxes,  make most of the above systems irreversible.

Stokes equation, $\nabla^2 {\bf u} = 0$, is often used to describe life at low-Reynolds numbers~\cite{Purcell:AJP1977}. Strangely,  Stokes equation is time reversible despite its dissipative nature. Hence, a swimmer cannot swim in such a flow by a \textit{reciprocal motion}, which is a statement of  \textit{scallop theorem}~\cite{Purcell:AJP1977}. To break the $T$ symmetry, researchers have proposed successful mechanisms including multiple hinges~\cite{Purcell:AJP1977} and inertia~\cite{Hubert:PRL2021,Lauga:RPP2009}. Here,  the $T$ symmetry can be broken by asymmetric time-dependent forces and asymmetric changes in body shapes; these mechanisms  are  not related to the second law of thermodynamics.

Chemical reactions in closed systems are time reversible, but they are typically irreversible in  open systems because of energy and particle exchanges between the system and environment.  For example, combustion   is an irreversible process, where hydrocarbons produce carbon dioxide, water,  heat, and sound.  

Biological organisms are supported by a large number of irreversible chemical and physical processes.
In human body, glucose is converted to energy via irreversible chemical process:
\be
\mathrm{C_6 H_{12} O_{6}}+ 6 \mathrm{O_2}\to  6 \mathrm{CO_2}  +6  \mathrm{H_2 O}.
\ee
Soft cartilages transform to bones via ossification, which is a irreversible process because a bone does  not revert back to a cartilage. Thus, irreversible chemical and biological processes break the $T$ symmetry in life forms. This is in addition to the role played by entropy and energy transfers.   \textit{Ageing} and \textit{decay} are related phenomena, but these topics are beyond the scope of this {   Perspective}.

\vspace{0.3cm} 
{\bf Time reversal symmetry vs.~other symmetry.}--
Fundamental laws of physics are symmetric  under  \textit{time translation, space translation}, and \textit{space rotation}, which leads to \textit{conservation of energy, linear momentum}, and \textit{angular momentum} respectively~\cite{Coleman:book:Aspects}. These conservation laws hold for isolated systems at all scales (from quarks to galaxies), but they  can be easily generalised to open systems by  accounting for the energy and momentum exchanges with the environment. The experiments to test  the above symmetries are relatively easy to design and perform~\cite{Feynman:book:Lectures1}.

In addition to the above three symmetries, there are three discrete symmetries---parity ($P$), charge conjugation ($C$), and time reversal ($T$)---that are respected by gravity, electromagnetism, and strong nuclear force, but violated minimally by the weak nuclear force~\cite{Feynman:book:Lectures1,Coleman:book:Aspects}.  For the microscopic physics, Wu~\cite{Wu:PR1957} 	reported $P$ and $C$ symmetry violations in \textit{beta decay}, whereas Christenson et al.~\cite{Cronin:PRL1964} reported $CP$  violation in \textit{neutral kaon decay}.  Later, it was shown that the $T$ symmetry is broken in  \textit{neutral K} system~\cite{Kane:book:Particle}. Interestingly, $CPT$ symmetry is respected by all the forces.  Hence, $CP$ violation implies $T$ violation. { Refer to Blum and Mart{\'\i}nez de Velasco~\cite{Blum:EPJH2022} for historical account of $CPT$  theorem.}

Note that the $P$ and $T$ symmetries are respected at microscopic level by gravity and electromagnetic forces. But, these symmetries are strongly broken at macroscopic level due to various factors. As discussed in this {   Perspective},   $T$ symmetry breaking at macroscopic scales is not related to  $T$ symmetry breaking  via the weak nuclear force.


Parity or mirror symmetry is relatively easy to test, both at microscopic and macroscopic scales. As stated by Feynman~\cite{Feynman:book:Lectures1}, parity operation involves changing left-helical components to right-helical ones, and vice versa.   The charge conjugation operation, which is exchanging particles with antiparticles, is not difficult either. Thus, we can easily perform mirror and charge conjugation at microscopic and macroscopic scales. 

However,   time reversal  operation on macroscopic systems is often impractical. For an organism, time reversal operation would include velocity reversals of the gases  being breathed in and out. Clearly,  inhaling carbon dioxide and exhaling oxygen  will kill the organism.  Reversal of blood flow would entail CO2-rich blood entering the cells, and O2-rich blood leaving the cells; another process that is injurious to the organism. In addition, a precise reversal of the velocities of billions and billions of molecules of an ideal gas is impractical (see Fig.~\ref{fig:Lebowitz}).  Breakage of a glass or an egg cannot be reversed because it  involves many irreversible physical processes, such as crack propagation in solids. Formation of large molecule  via atomic collisions cannot be reversed  because of inherent quantum interactions.  These inherent difficulties in performing time-reversals operation appear to set apart the $T$ symmetry from other symmetries.  I believe that we need to examine this issue carefully.


Another factor that makes time reversal symmetry unique is \textit{causality}~\cite{Causality:book}, which is an important assumption of physics. If event $A$ is caused by an event $B$, then event $A$ occurs after event $B$. For example, a light bulb lights up (an effect) after  its switch is turned on (a cause).  This fundamental principle breaks the time reversal symmetry in an apparently trivial way.  

Causality often involves an agent, which could be a human, nature, or some internal process. For example, breakage of an egg may involve a cook making an omelette, or an accidental drop of the egg on the floor. The experiment of Fig.~\ref{fig:Lebowitz} requires some agent to remove the separating wall between the two chambers.  We often ignore the role of an agent in \textit{objective description of physical phenomena}.  However, these agents play a critical role in natural world, especially for the arrow of time.

Like $T$ symmetry,  $P$ symmetry  is violated  at macroscopic scales. For example, human hearts resides in the left side of the body; sea shells are helical; many biomolecules, including DNA, are helical; many  fruits and plants exhibit helicity.  We may infer that the $P$ symmetry violation via weak nuclear force is not responsible for the asymmetry at the macroscopic level.  The common factors that   violate $P$ and $T$ symmetries are asymmetric chemical and evolutionary processes, and interactions with environment (e.g., interactions with the atmosphere and the Sun). The  factors involving $T$ symmetry breaking, but not $P$ symmetry breaking, are energy transfers and causality.   We do not  dig deeper into $P$ symmetry breaking   because it is beyond the scope of this {   Perspective}.

\vspace{0.3cm} 
{\bf Summary.}-- Fundamental laws of physics with gravity and electromagnetic forces respect the time reversal ($T$) symmetry. \textit{However,  the $T$ symmetry is broken very strongly in the macroscopic world}.   In this {   Perspective}, we discuss the frameworks of entropy, asymmetric energy transfers, and open systems that are invoked to break the $T$ symmetry macroscopically. In summary,

\begin{enumerate}
	\item The second law of thermodynamics (increase in Boltzmann entropy with time) successfully explains $T$ symmetry breaking for many macroscopic systems. But, the Boltzmann entropy is not well defined for many time irreversible systems---Euler and hydrodynamic turbulence, earthquakes, human body---for which we invoke the following frameworks for breaking the $T$ symmetry.
	
	\item Many multiscale systems are forced at large scales, and they have dissipation at small scales. In such systems, the forward energy flux  from large scales to small scales determines the arrow of time. In addition, the energy transfer is a good quantifier for $T$ symmetry breaking in energy-conserving systems, such as Euler turbulence, quantum turbulence, and molecular dynamics.   
	
	\item Open systems are time irreversible due to external forces and due to energy and particle  exchanges  with the environment.  For example, combustion is irreversible due to the  heat, sound,  and matter exchanges with the environment. The human body has a definite arrow of time due to many cellular and chemical reactions.  In open macroscopic world, \textit{Causality} is  another important factor  that breaks  $T$ symmetry.
	
	
\end{enumerate}

Boltzmann entropy is not clearly defined for many systems, e.g., turbulent flows, biological systems, earthquakes, etc.  Hence, to break the $T$ symmetry, it is prudent to employ the above factors that do not require explicit computation of entropy.  We remark that deductions based on idealizations---point particles, spherical balls, isolated systems---may lead to erroneous conclusions. Instead, simple arguments based on asymmetric energy transfers, open systems, causality etc., easily provide arrow of time for the natural world.

Fundamental and universal laws are useful in physics due to their wide scope. Still, many complex phenomena, especially in nonequilbrium world and biology, go  beyond the presently-available universal laws~\cite{Verma:IASCS2019}.  In this spirit, the present {   Perspective} attempts to explain $T$ symmetry breaking in the macroscopic world using factors beyond the second law of thermodynamics.  


***

The author thanks Anurag Gupta, Adhip Agarwala, Arvind Ayyer, Rodion Stepanov, and Frank Plunian for useful discussions. { The author thanks the referees and the  Editor-in-Chief for useful suggestions.} This work is supported by Science and Engineering Research Board, India (Grant numbers: SERB/PHY/20215225 and SERB/PHY/2021473), and J. C. Bose Fellowship (SERB /PHY/2023488).


\end{document}